\documentclass[12pt]{article}
\usepackage{epsfig}
\newcommand{\la}[1]{\label{#1}}
\newcommand{\eq}[1]{eq.~(\ref{#1})}

\newcommand{\Eq}[1]{Eq.~(\ref{#1})}

\newcommand{\half}{\frac{1}{2}}
  \def\beq{\begin{equation}}
  \def\eeq{\end{equation}}
  \def\beqr{\begin{eqnarray}}
  \def\eeqr{\end{eqnarray}}
  
 \def\Tr{\mbox{Tr}}

 \def\bra#1{{\langle#1\vert}}
 \def\ket#1{{\vert#1\rangle}}
 
 \def\Dirac#1{#1\hskip-6pt/}
 \def\dd{\Dirac\partial}

\begin{document}
\thispagestyle{empty}
\begin{flushleft}
\vspace*{1cm}
\rightline{ PNPI--2500}
\rightline{ RUB-TP2-19/02}
\rightline{December 2002}

\vspace*{2cm}
{\Large \bf Light cone nucleon wave function

in the quark-soliton model}\\
\bigskip

V. Yu.  Petrov$^\diamond$ and M. V. Polyakov$^\diamond$$^*$\\

\vspace*{1cm}
$^\diamond$
Petersburg Nuclear Physics Institute, Gatchina, 188300, Russia.\\
$^*$Institute for Theoretical Physics II, Ruhr University, 4470 Bochum,
Germany.
\end{flushleft}

\begin{center}
\bf A b s t r a c t
\end{center}

The light-cone wave function of the nucleon is calculated in the limit
$N_c\to\infty$ in the quark-soliton model inspired by
the theory of the instanton vacuum of
QCD. The technique of the finite time evolution operator is used in order
to derive expressions for all components of the Fock vector describing the nucleon in the
infinite momentum frame. It is shown that nucleon wave function for large
$N_c$ can be expressed in terms of the wave function of the discrete
level in the self-consistent meson field and light cone wave functions
of 1,2, etc mesons. The 3-quark components of the nucleon and
$\Delta$-resonance are estimated. Wave function of the nucleon appears
to be positive in the whole region of $x$ and it has rather small
asymmetry. It differs strongly both from Chernyak-Zhitnitsky wave
function and the asymptotic one. Large momentum transfer asymptotic of
the electromagnetic and axial
form factors is discussed.

\section{Introduction}

Light cone wave functions of hadrons were introduced in
hadron physics many years ago \cite{Chernyak77,Brodsky79,Efremov79,Brodsky98,Lepage80}.
They contain virtually the
full information which is necessary to describe hadron properties
at high energies. The use of hadron wave functions in the context in
QCD relies on the concept of factorization. Processes with hadrons
are divided into 2 parts: i) the hard process-dependent parts are
calculated according to perturbative QCD and ii) the soft
process-independent part is usually encoded in soft functions,
parton distributions, fragmentation functions, etc. Usually, a formal
definition of the soft part can be formulated in terms of definite
quark and gluon hadron matrix elements. Their logarithmic scale
dependence is well understood in terms of corresponding evolution
equations.

In principle, any soft part can be expressed in terms of the
hadron wave function. For example, ordinary parton distributions
is a sum of wave functions squared corresponding to a different number
of quarks or gluons integrated over all momenta of partons except one.
The analogous definition in terms of wave functions can be done for
the fragmentation functions as well.

Another type of hadronic matrix elements is involved
in the description of the elastic form
factors or transition form factors. There one deals with the matrix elements of quark
currents between hadron states which have very different initial and
final momenta.  The large difference of momenta in exclusive reactions
effectively separates out the component of wave function with the
minimal number of constituents, so gives an access to the simplest
structures in hadron \cite{Chernyak77,Brodsky79,Efremov79,Brodsky98,Lepage80}.

An intermediate situation between these two types of reactions
occurs in processes like deeply virtual Compton scattering and
hard meson production. In this case, the hadron operators involved are
bilocal and their matrix elements are off-forward.
The soft part of such processes is described by so-called
generalized distributions. Recently  it was shown that
generalized distributions can also be easily presented in terms of light
cone wave functions \cite{Jacob}.

Unfortunately up to now light-cone wave functions of baryons (as well
as mesons) in the low normalization point cannot be determined from the
first principles of QCD. Perturbative theory is able to predict
only so-called asymptotic wave functions which are normalized at
arbitrary high normalization point.

Up to the present moment the light-cone nucleon wave functions were
calculated only in the framework of QCD sum rules \cite{Chernyak}. Of
course, this
can be considered only as a crude estimate.  For this reason
self-consistent models of the nucleon (more or less motivated by QCD)
become highly desirable.  Unfortunately, to the best of our knowledge,
no calculation of the nucleon wave function was made in any
self-consistent relativistic field-theoretical model.

In this paper we attempt to calculate light-cone wave functions at a
low normalization point in the limit of large number of colours, $N_c
\rightarrow\infty$. Even though in reality $N_c=3$, the academic limit
of large $N_c$ is known to be a useful guideline. It is a general
QCD theorem that at large $N_c$ the nucleon is heavy and can be viewed
as a classical soliton \cite{Witten83}.

An example of the dynamical realization of this idea is  given by the
Skyrme model \cite{Adkins83}. However, the Skyrme model is based on
an unrealistic effective chiral Lagrangian.
A far more realistic effective chiral Lagrangian is given by the
functional integral over quarks in the background pion field
\cite{Diakonov86}:
\[
\exp\left(iS_{\rm eff}[\pi(x)]\right)=
\int D\psi D\bar\psi \exp\left(i\int d^4x
\bar\psi(i\dd - MU^{\gamma_5})\psi\right),
\]
\beq
U=\exp\left(i\pi^a(x)\tau^a\right),\;\;\;\;\;
U^{\gamma_5}=\exp\left(i\pi^a(x)\tau^a\gamma_5\right)=
\frac{1+\gamma_5}2 U
+ \frac{1-\gamma_5}2 U^\dagger.
\la{FI}\eeq
Here $\psi$ is the quark field, $M$ is the effective quark mass
which is due to the spontaneous breakdown of chiral symmetry (generally
speaking, it is momentum-dependent) and $U$ is the $SU(2)$ chiral pion
field. The NJL-type Lagrangian \Eq{FI} has been derived from the
instanton model of the QCD vacuum \cite{DP1,DPP},
which provides a natural mechanism of chiral symmetry breaking and
enables one to express the dynamical mass
$M$ and the ultraviolet cutoff $\Lambda$ intrinsic to \eq{FI} through
the $\Lambda_{QCD}$ parameter. Proposed on the basis of Lagrangian of
\eq{FI} the chiral quark-soliton model~\cite{DPP,DPP2,DPP3,DPP4} describes
properties of baryons far better than the Skyrme model.  For the recent
status of the chiral quark-soliton model see reviews \cite{Rev1,Rev2}.

Recently in the framework of quark-soliton the
parton distributions
\cite{Part1,Part2} and the generalized distributions~\cite{Offw}
have been computed
at a low normalization point. In
this paper we complete the program of the investigation of high-energy
properties of baryons in the quark-soliton model and present the
framework for calculation of the light-cone wave of the nucleon.

We use the evolution operator technique in order to present wave
functions of the nucleon state moving with the speed $V$.
In the limit $V\rightarrow 1$ the corresponding wave function tends
to the light-cone wave function in question. This method is convenient
as the evolution operator can immediately be  expressed as
a functional integral with definite boundary conditions.

At large $N_c$ the functional integral representing the evolution
operator can be evaluated using the saddle-point method.
The saddle-point of the effective Lagrangian in \eq{FI} in the sector
with unity baryon charge corresponds to the
nucleon-soliton \cite{DPP3}.  The stationary solutions of saddle-point
equations correspond to the nucleon in the rest frame. Owing to
relativistic invariance of the equations of motion there is also
infinite number of other solutions which describe the moving solitons.
We need the solution which describes the nucleon in the infinite
momentum frame, and we have to extract the corresponding
quark-antiquark wave function. In the $N_c\to\infty$ limit  it can be
viewed as a product of the quark states in the time-dependent
self-consistent pion field.

We show that the nucleon wave function at large $N_c$ is a product
of $N_c$ one-quark wave functions of the {\em valence quarks}
and the coherent exponential of quark-antiquark pairs corresponding
to the {\em sea quarks}. As it should be in this limit, the nucleon wave
function is completely factorized in colours. One-quark wave functions
receive contribution both from the wave function of the {\em discrete
level} in the mean pion field and the sea quarks. The quark-antiquark
pair wave function can be expressed in terms of so-called Feynman Green
function at finite time.

This structure of the light-cone nucleon wave function is rather
general. Indeed, as was already said, nucleon is
a soliton of an effective meson Lagrangian at large $N_c$. Hence
its quark wave function is a product of states in the external field of all
mesons. The model with Lagrangian of \eq{FI} (which is, in turn, based
on the theory of the instanton vacuum) is specific in two respects. First, as the
size of the nucleon ($\sim 1/M$) is parametrically large as compared to
the ordinary hadron scale ($\sim 1/\rho$, $\rho$ being the size of the
instanton), only the lightest degrees of freedom (i.e. pions and
constituent quarks) are important. Second, the number of {\em gluons}
is suppressed in the instanton vacuum by the parameter $(M\rho)^2\ll
1$ \cite{Part1,Rev2}, so gluons in this model do not participate in the
formation of the nucleon wave function.

All what one needs to know in order to calculate the nucleon wave
function in the model is the wave function of the discrete level (the
solution of the Dirac equation in the external field) and the
quark-antiquark pair wave function. Expanding the latter wave function
in $\pi$-meson field, it is possible to present the wave function
as  convolution of
the pion mean field with the quark-antiquark light-cone wave function
of 1,2,$\ldots$ pions. This is to be expected, as the pions are the only
agents inducing the interaction in the model.

In fact, light-cone wave functions of  one and two pions were
already considered in the instanton vacuum  (see Refs.~\cite{pionwf}
and \cite{maxchrist}, correspondingly) and appeared to describe the data
correctly.  We discuss the connection with meson wave functions in
the Section~3.

We formulate the scheme for calculation of the
nucleon-soliton light-cone wave function in Section~2 and
in Section~3 we consider the wave functions in the infinite momentum frame. The three-quark
components of the nucleon wave functions (so-called {\em distribution
amplitude}) is calculated in Section~4. They appear to be almost
symmetric (antisymmetric part of the nucleon wave function is
numerically small) smooth functions, which are far  from both
the asymptotical wave function and the wave function of
Chernyak-Zhitnicky type~\cite{Chernyak84a,Chernyak84b}.

We also discuss  physical observables
(asymptotics of electromagnetic and axial form factors) and evolution
of the model wave function to the high $Q^2$. Asymptotics of the
form factors cannot be calculated explicitly as corresponding integrals
depend strongly on the region of small $x$ where the model is not
valid. However we are able to calculate {\em ratios} of the form factors
which appear to be rather close to the experimental data.

\section{Soliton wave function in field theory \newline at
large~$N_c$}

In principle, calculation of the wave function of a given state
in terms of quarks and antiquarks is straightforward in the quantum
field theory. However, usually this task is too complicated. In fact, we
know only one model field theory, namely, the Schwinger model where
the program of calculating of all wave functions was
completed~\cite{Danilov80}. The most direct way to obtain wave
functions of any state is to calculate the evolution operator $S(T)$
for the given theory and present it as a sum:
\beq
S(T)=\exp(-i \hat{H} T)=\sum_{n}e^{-iE_nT}\ket{n}\bra{n}.
\la{evolution-operator}
\eeq
Here $|n\rangle$ is vector of certain state ---
the eigenfunction of Hamiltonian $H$. As it is well-known, the
evolution operator of \eq{evolution-operator} can be expressed as a
functional integral at {\em finite time} with definite boundary
conditions, namely \footnote{We immediately write down the  functional
integral in the model of \eq{FI} but, in fact, we have to start from
the full QCD. The corresponding functional integral is the integral
over both the quark and gluon field. In the model of the instanton
vacuum this integral is saturated by instanton and antiinstanton gluon
configuration.  One can then integrate out gluons in this
approximation. As it was shown in Refs.~\cite{DP1,DPP}, one obtains
the low-energy effective Lagrangian \eq{FI}. As the number of gluons in the
nucleon in the instanton model is parametrically suppressed, we should
not put any boundary conditions on gluon field. For this reason the
derivation of the evolution operator in the low-energy limit repeats
literally the derivation of the low-energy effective lagrangian
that was done in these papers. If one is still  interested in gluon
components of the wave function, they should be traced directly from
the general expression for the full QCD evolution operator to which we
have to apply the same approximations which lead to the effective
Lagrangian of \eq{FI}. This is straightforward in the instanton vacuum
model.  Gluon components are of the order of $(M\rho)^2\ll 1$ where
$\rho$ is the instanton size. The same parameter was used in the
derivation of effective lagrangian \eq{FI}}

\beq
S[T]=\int_{\psi^{(+)}=a}^{\psi^{(-)}=b^+}\!\!D\psi(x)
\int_{\bar{\psi}^{(+)}=b}^{\bar{\psi}^{(-)}=a^+}\!\!
D\bar{\psi}(x)\int\! D\pi(x)
\exp\left(i\!\int^T_0\!\! dt\;{\cal L}_{eff}\right),
\la{functional-integral}
\eeq
where ${\cal L}_{eff}$ is the effective Lagrangian of \eq{FI}.
The operators $a^\pm(p)$,  $b^\pm(p)$ are annihilation-creation
operators for quarks-antiquarks which are defined through the expansion
of the field $\psi(x)$ into positive- and negative-frequency parts:
\beq
\psi_\alpha(x)=\int\!\frac{d^3p}{(2\pi)^3}\sum_\lambda\sqrt{\frac{m}{p_0}}
\left(
a^{(\lambda)}(\vec{p})u^{(\lambda)}(\vec{p})e^{i\vec{p}\cdot\vec{x}}
+b^{(\lambda)+}(\vec{p})v^{(\lambda)}(\vec{p})e^{-i\vec{p}\cdot\vec{x}}
\right)
\la{creation-annihilation}
\eeq
(here $\lambda$ is quark polarization;
$u^{(\lambda)}(\vec{p})$, $v^{(\lambda)}(\vec{p})$
are free plane wave spinors).

We do not impose any boundary conditions on the $\pi$-meson field. This
field appears in the derivation of the effective Lagrangian as a result
of bosonization~\cite{Diakonov86}, and it should not be considered as
an elementary one. In fact, it is impossible to consider the light-cone
wave function in terms of {\em both} quarks and
$\pi$-mesons --- it would be a kind of double-counting.

In the representation of \eq{functional-integral} the quark-antiquark
creation operators $a^+,b^+ $ anticommute with annihilation
operators $a,b$, as they are related to different times $0$ and $T$.
One has to calculate the functional integral of \eq{functional-integral}
at the {\em finite} time $T$ as a functional of operators
$a^\pm$, $b^\pm$. Expanding in the exponentials $\exp(-iE_nT)$ and
factorizing creation operators
from annihilation ones, we can find eigenfunctions of all
states in terms of quarks-antiquarks (see, e.g.~\cite{Danilov80}).

In the large $N_c$-limit the functional integral over the $\pi$-meson
field should be calculated in the saddle-point approximation. This
procedure can be formulated as follows.

Let us first integrate  over the quark field $\psi(x)$. We divide
fermion field into two parts, $\tilde{\psi}(x)$ and $\chi(x)$:  \beq
\psi(x)=\tilde{\psi}(x)+\chi(x), \quad
\bar{\psi}(x)=\bar{\tilde{\psi}}(x)+\bar{\chi}(x).
\la{sum-of-terms}
\eeq
Let us require that $\tilde{\psi}(x)$ obeys the Dirac equation in the
external $\pi$-meson field with corresponding boundary conditions:
\[
\left(i\hat{\partial}-Me^{i\hat{\pi}(x)\gamma_5}
\right)\tilde{\psi}=0,
\]
\[
\tilde{\psi}^{(+)}(\vec{x},t=0)=
A(\vec{x})\equiv
\int\!\frac{d^3p}{(2\pi)^3}e^{i\vec{p}\cdot\vec{x}}
\sqrt{\frac{p_0}{m}}\sum_\lambda
a_{\lambda}(\vec{p})u^\lambda(\vec{p}),
\]
\beq
\tilde{\psi}^{(-)}(\vec{x},t=T)=
B^+(x)\equiv
\int\!\frac{d^3p}{(2\pi)^3}e^{-i\vec{p}\cdot\vec{x}}
\sqrt{\frac{p_0}{m}}\sum_\lambda
b_\lambda^+(\vec{p})v^\lambda(\vec{p}),
\la{Dirac-eq} \eeq
The field $\chi(x)$ obeys zero boundary conditions.

A solution of the Dirac \eq{Dirac-eq} can be expressed in terms of
the {\em finite time} Feynman Green function
$G^{(T)}(\vec{x},t|\vec{y},0)$
\[
\tilde{\psi}(x,t)=\int d^3y
\int\frac{d^3p}{(2\pi^3)}\sqrt{\frac{p_0}{M}}\sum_\lambda
\left[
G^{(T)}(\vec{x},t|\vec{y},0)u^{\lambda}(\vec{p})a_\lambda(\vec{p})
e^{i\vec{p}\cdot\vec{y}}
\right.+
\]\beq
\left.
+G^{(T)}(\vec{x},t|\vec{y},T)v^{\lambda}(\vec{p})b_\lambda^+(\vec{p})
e^{-i\vec{p}\cdot\vec{y}}
\right],
\la{Dirac-solution}
\eeq
with an analogous expression for $\bar{\tilde\psi}(\vec{x},t)$.

The finite time Green function
$G^{(T)}(\vec{x},t|\vec{x}^\prime,t^\prime)$ is the solution of the
Dirac equation in the external field which has vanishing
negative-frequency part at $t=0$ and vanishing positive-frequency part
at $t=T$. At $T\to\infty$ finite time Green function reduces to the
usual Feynman Green function. Feynman propagators for free particles
appear to obey automatically the required boundary conditions at any
$T$.  Also it is easy to verify that the finite time Green function in
the external field can be constructed as a sum of diagrams with free
Feynman propagator interacting with external field.  The only
difference with the usual Feynman Green function is that all integrals
in intermediate points should be carried only in the finite time
interval $0<t<T$ (for Feynman Green function one has to integrate over
the whole interval $-\infty<t<\infty$).

Let us substitute the expansion of \eq{sum-of-terms} into the
functional integral of \eq{functional-integral}.
One can  see that variables $\tilde{\psi}$ and $\chi(x)$ are completely
separated from each other. As a result we arrive at
the following expression for the evolution operator (cf
\cite{Danilov80}):
\beq S[T,a^\pm,b^\pm]=\int\! D\pi(x)
{\rm Det}^{(T)}[\pi]S_{ext}[\pi,T]
\eeq
(we used here \eq{Dirac-solution} for $\tilde{\psi}$).
Here $S_{ext}[\pi,T]$ is the evolution operator in
the given external field:
\[ S_{ext}[\pi,T]=\exp\!\left\{\int\!
d^3xd^3y\!
\left[ B(x)\gamma^0G^{(T)}(x0,yT)B^+(y)+
\right.
\right.
\]\[
+B(x)\gamma^0G^{(T)}(x\varepsilon,y0)A(y)
+A^+(x)\gamma^0G^{(T)}(xT,y0)A(y)+
\]\beq
 \left.\left.
+A^+(x)\gamma^0G^{(T)}(xT-\varepsilon,yT)B^+(y)
\right]
\right\},
\la{pi-integral}
\eeq
where $\varepsilon\to +0$); the quantity $ {\rm Det}^{(T)}[\pi]$ is the
{\em finite-time determinant} in the
external field. It is the Gaussian functional integral over the fermion
field $\chi(x)$ with zero boundary conditions. For this reason it does
not depend on the operators $a^\pm$, $b^\pm$ being only the functional
of pion field $\pi(x)$. It can also be presented in terms of the
finite-time Green function:
\beq
{\rm Det}^{(T)}[\pi]=\exp\left[\int_0^M\!
dM \int_0^T\!  dt\int\! d^3x \Tr \left(G^{(T)}(\vec{x} t,\vec{x}t)i
U^{\gamma_5}(\vec{x},t)
\right) \right]
.
\la{det-green}
\eeq

Thus the evolution operator in the external field is the coherent
exponential of the creation-annihilation operators. The nucleon is the
lowest possible state in the sector with baryon charge $B=1$. One can
obtain its wave function by applying the evolution operator to any
colourless state of $N_c$ quarks and taking the  $T\to\infty$
limit. For example, the nucleon wave function $\Phi_N$ can be obtained
from the state of free quarks:
\[ c_{N}e^{-i{\cal M}_N T}\Phi_N(a^+,b^+)=
\lim_{T\to\infty}S(T)\prod_{i=1}^{N_c}
a^+_{\alpha_i}(p_i)\ket{\Omega_0}\sim
\]
\beq
\sim \int\!\! D\pi {\rm Det}^{(T)}[\pi] \prod_i^{N_c}\!
G_i^{(T)}(p_i,0,k_i,T)a^+_i(k)
\exp[a^+(p)\ G^{(T)}(T,p,T,p^\prime)\ b^+(p^\prime)]
\la{nwf-origin}
\eeq
(we
write the formula in a bit symbolical form, at the moment it is
sufficient for our purposes). Here ${\cal M}_N$ is the nucleon mass
and $c_{N}$ is the overlap between the initial state of $N_c$ quarks
and nucleon wave function. The functional integral of \eq{nwf-origin}
should be calculated in the saddle-point approximation. One has to find
the pion field which extremizes the integrand. At large $T$ important
factors are \footnote{Operator exponential in \eq{nwf-origin} does not
contribute to the saddle-point \eq{full-energy} as the Green function
$G^{(T)}(T,p,T,p^\prime)$ does not contain exponential with the phase
proportional to the time $T$. Let us note also that minimizing of
\eq{full-energy} gives the saddle-point for pion field only at times
which are far from the end-points of the interval $(0,T)$. In fact this
is enough for the calculation of wave function. }
\beq
{\rm Det}^{(T)}[\pi]\sim \exp(-iE_{field}[\pi]T),
\quad
G_i^{(T)}(p_i,0,k_i,T) \sim
\exp(-iE_{level}[\pi]T),
\eeq
where $E_{field}[\pi]$ is the energy of the Dirac continuum with the
given pion field (proportional to $N_c$) and $E_{level}[\pi]$
is the energy of the (possible) discrete level for the quark in this
field. In order to find the saddle-point one has to minimize the sum:
\beq
{\cal E}[\pi]=E_{field}[\pi]+N_c E_{level}[\pi]
\la{full-energy}
\eeq
in the presence of the pion field. It is exactly the condition which
was used in constructing the nucleon-soliton in
Refs~\cite{DPP,DPP2,DPP3,DPP4}.  Both contributions to the total energy
are of the order of $N_c$. The first is the full effective chiral
Lagrangian (ECL) in the low-energy effective theory of \eq{FI}
calculated for a  given pion field $\pi$.

The minimum of \eq{full-energy} is achieved at some stationary
pion field and corresponds to the nucleon at rest. The value of
the energy in the minimum is the nucleon mass:
\beq
{\cal M}_N=min\; {\cal E}[\pi] \sim O(N_c).
\eeq
This minimum was found in Ref.~\cite{DPP3,DPP4} and it corresponds to
the hedgehog symmetry of the pion field:
\beq
\bar{\pi}^a(\vec{x})=n^a
P(r), \quad \vec{n}=\frac{r}{r},
\la{hedgehog}
\eeq
where the profile function $P(r)$ is to be calculated numerically.

We can obtain the wave function of nucleon in the leading order of
$N_c$, if we substitute the saddle-point field $\pi(x)$ of
\eq{hedgehog} into \eq{nwf-origin}. In the higher orders, one has to
express the general pion field $\pi(x,t)=\bar{\pi}+\pi^{quant}$
and then perform the Gaussian integration in $\pi^{quant}$ which
appears to be $\pi^{quant}\sim O(1/\sqrt{N_c})$. Thus it is possible
to formulate the systematic perturbation theory in $1/N_c$. However, in
this paper, we restrict ourselves to the leading order.

Let us stress that the procedure of the calculation of the nucleon
wave function which we have formulated, is rather general. Indeed, it
is a general QCD theorem that at large $N_c$, the nucleon is the soliton
of some effective meson Lagrangian. Thus its wave function can always
be  presented in the form of \eq{nwf-origin} where Green functions
should be found in the self-consistent field of all mesons entering
this effective Lagrangian. Of course, the exact low-energy
meson Lagrangian is unknown. In the present work, we use  the
instanton vacuum model in order to fix this low-energy Lagrangian.

Let us also rewrite nucleon wave function in a different form.
As explained above, nucleon can be described as $N_c$ valence quarks
+ Dirac continuum in the self-consistent external field. It is clear
from \eq{nwf-origin} that wave function of the Dirac continuum
(i.e. the state with all states with negative-energies occupied)
is the coherent exponential of the quark-antiquark pairs:
\[
\ket{\Omega}=\exp\left[\sum_{colour}\int d^3xd^3y\
A^+(x)\ \gamma^0G^{(T)}(xT-\varepsilon,yT)\ B^+(y)
\right]\ket{\Omega_0}\equiv
\]
\beq
\equiv
\exp\left[\sum_{colour}
\int\!\frac{d^3p_1d^3p_2}{(2\pi)^3}\ a^+_{\lambda_1}(\vec{p}_1)\
\Theta^{\lambda_1,\lambda_2}(\vec{p}_1,\vec{p}_2)\ b^+_{\lambda_2}(\vec{p}_1)
\right]\ket{\Omega_0},
\la{wf-pairs}
\eeq
where $\ket{\Omega_0}$ is the vacuum of quarks-antiquarks. The function
$\Theta^{\lambda_1,\lambda_2}(\vec{p}_1,\vec{p}_2)$ can be called
the wave function of the quark-antiquark pair. We see that it is
expressed in terms of the finite time Green function at equal times.

The nucleon itself is a state with one more level  occupied by
valence quarks, namely, a discrete level (with positive
energy) which appears in the self-consistent external field.
We can obtain the nucleon wave function by applying
the operator which fills this discrete level to \eq{wf-pairs}:
\beq
\Phi_N=\prod_{colour}\int d^3x \psi^+(x) f_{\rm lev}(x)\ket{\Omega},
\eeq
where $f_{\rm lev}(x)$ is the wave function of the discrete level (solution
of Dirac equation in the external $\pi$-meson field). In order to
express $\Phi_N$ in terms of quark-antiquark we have to expand
$\psi$-operators according to \eq{creation-annihilation} and commute
them with the exponential of \eq{wf-pairs}. As a result we get the
following expression for the nucleon wave function ($A$ is the
normalizing constant):
\[
\Phi_N=A\prod_{colour}\int
\frac{d^3p}{(2\pi)^3}\ F^{\lambda}(\vec{p})\ a^+_\lambda(\vec{p})
\times
\]\beq
\times\exp\left[\sum_{colour}
\int\!\frac{d^3p_1d^3p_2}{(2\pi)^3}\ a^+_{\lambda_1}(\vec{p}_1)\
\Theta^{\lambda_1,\lambda_2}(\vec{p}_1,\vec{p}_2)\ b^+_{\lambda_2}(\vec{p}_1)
\right]
|\Omega_0\rangle
\la{wf-total}
\eeq
Here $F^{\lambda}(\vec{p})$ is the one-quark wave function. It is a
sum of two contributions:
\[
F^\lambda(\vec{p})=\int\frac{d^3p^\prime}{(2\pi)^3}\sqrt{\frac{M}{
\omega^\prime}}
\left[
u^{*\lambda}(\vec{p})\ f_{\rm lev}(\vec{p})\ (2\pi)^3\delta^{(3)}\left(\vec{p}-
\vec{p}^\prime\right)-
\right.\]\beq
\left. -
\Theta^{\lambda,\lambda^\prime}(\vec{p},\vec{p}^\prime)\
v^{*\lambda^\prime}(\vec{p}^\prime)\
f_{\rm lev}(-\vec{p}^\prime)
\right]
\la{1-quark}
\eeq
The first contribution in \eq{1-quark} is that of valence
quarks, while the second term can be called the contribution of the sea
quarks to one-quark wave function.

Let us point out  the fact that the nucleon wave function is
completely factorized in colours. In fact, it is a general theorem in
the strict $N_c\to\infty$
limit.

\section{Wave function of the nucleon in the IMF}

It is well-known that the wave function in the rest frame has not too much
physical sense \cite{Lepage80}. The physical meaning can be ascribed
only to the wave function of the fastly moving nucleon (nucleon in the
infinite momentum frame). Contrary to the wave function in the rest
frame it can be accessed by measurements.

The common approach to the hadron wave functions is the formalism
based on the light-cone quantization (see, e.g., review~\cite{LC-Ham}).
This formalism has many obvious advantages as compared to the approach
based on Schr\"odinger wave functions. From the
other side, the Schr\"odinger equal-time wave function can be defined in
any frame, not necessarily on the light-cone. Also one can use
well-trodden path which starts from the calculation of the evolution
operator in the functional integral technique. This is why we use in
this paper the last method. Light-cone wave function of the nucleon is,
by definition, its wave function in the {\em infinite momentum frame}
\cite{Lepage80,Lepage79b}.

The stationary saddle-point $\bar{\pi}^a(\vec{x})$ corresponds to the
nucleon-soliton at rest. However, as the effective chiral Lagrangian is
relativistically invariant, we are guaranteed that there are infinitely
many solutions of saddle-point equations of motion which describe the
nucleon moving in some direction with a speed $\vec{V}$. The
corresponding pion field is time-dependent and can be obtained from
the stationary field by Lorentz transformation
\beq
\pi^{(cl)}(\vec{x}, t)=\bar{\pi}\left(
\frac{x-\vec{V}t}{\sqrt{1-V^2}}
\right)
\la{pi-moving}.
\eeq
In order to find all states of the Dirac continuum in the moving
nucleon it is sufficient to solve the Dirac equation in the field of
\eq{pi-moving}. In particular, the wave function of the valence level
can also be obtained as Lorentz-transformation:
\beq
\Phi_{\rm lev}(\vec{x},t)=S[V]f_{\rm lev}(x)\left(\frac{\vec{x}-\vec{V}t}
{\sqrt{1-V^2}}\right)\exp\left( -i
\frac{\varepsilon t+\vec{V}\vec{x}}{\sqrt{1-V^2}}
\right),
\la{level-moving}
\eeq
where $\varepsilon$ is the energy of the discrete level. Here $S[V]$ is a
matrix which transforms Lorentz indices
\beq
S\left[V\right]=\exp(i\sigma_{03}\omega), \quad \sigma_{\mu\nu}=\frac{1}{2}
[\gamma_\mu,\gamma_\nu],\quad {\rm tanh}(\omega)=V.
\la{matS}
\eeq

The wave function of the discrete level in stationary field
$\bar{\pi}(\vec{x})$ of \eq{hedgehog} is a mixture of two wave
functions with orbital angular momentum $L=0$, $h(r)$,
and $L=1$, $j(r)$ \cite{DPP2,DPP3}:
\beq
f^{\alpha f}_{\rm lev}(\vec{x})=\frac{1}{4\sqrt{\pi}}
\left(
\begin{array}{c}
-h(r)\varepsilon^{\alpha
f}+ij(r)\frac{\vec{r}}{r}\vec{\sigma}^\alpha_\beta\varepsilon^{\beta
f}\\
h(r)\varepsilon^{\alpha
f}+ij(r)\frac{\vec{r}}{r}\vec{\sigma}^\alpha_\beta\varepsilon^{\beta
f}\\
\end{array}
\right).
\la{level-wf}
\eeq
Here $\alpha$ is a spinor index and $f$ is flavour index.
The spherically symmetric functions $f(r)$ and $g(r)$ were found by
numerical integration of the Dirac equation.

Let us begin with the calculation of the contribution of the discrete level
in the one-quark wave function (first term in \eq{1-quark}) in the infinite
momentum frame ($V\to 1$). We proceed in \eq{level-moving} to momentum
space and obtain from \eq{1-quark}:
\[
F^{\lambda f}_{\rm lev}(p)=
\sqrt{\frac{m}{\omega_p}}
\int d^3k\
u^{*(\lambda)}_\alpha(p)\
S[V]^{\alpha}_{\beta}\
\tilde{f}^{\beta f}_{\rm lev}
(\vec{k})\
\delta^{(2)}\left(p_\perp-k_\perp\right)\delta\left(\frac{k_3+\varepsilon}{\sqrt{1-V^2}}-p_3\right),
\]
\beq
\tilde{f}^{\alpha f}_{\rm lev}(\vec{k})
=\int d^3x\ e^{-i\vec{k}\cdot\vec{x}}
f^{\alpha f}_{\rm lev}(\vec{x}).
\la{tmp1}
\eeq
Let us divide the quark momentum into the longitudinal and transverse
parts with respect to the total nucleon momentum
$P_N={\cal M}_NV/\sqrt{1-V^2}$:
\[
p=\left(zP_N,\vec{p}_\perp
\right)
.
\]
Also in
the IMF it is convenient to use quark-antiquark operators normalized in
a different way than the usual ones:
\beq
\{\bar{a}^+(z_1,\vec{p}_{1\perp}),
\bar{a}(z_2,\vec{p}_{2\perp})
\}=\delta
\left({z_1-z_2}
\right)(2\pi)^2\delta^{(2)}\left(
\vec{p}_{1\perp}-\vec{p}_{2\perp}
\right)
.
\la{quarkops}
\eeq
(We shall call a wave function the coefficient in front of quark operators
normalized this way). We have from \eq{tmp1}
\beq
\bar{F}^{\lambda f}_{\rm lev}(z,\vec{p}_\perp)=
\frac{{\cal
M}_N}{P_N}\sqrt\frac{M}{z}\bar{u}(\vec{p})S[V]\tilde{f}_{\rm lev}
(\vec{p})|_{p_3=z{\cal M}_N-\varepsilon}
.
\eeq
We make use of the relation $\gamma^0S[V]=S^{-1}[V]\gamma^0$ and apply
Lorentz transformation to the free spinor $\bar{u}(p)$:
\beq
\bar{u}^{(\lambda)}(p)S^{-1}[V]=\bar{u}^{\lambda}(\tilde{p}),
\quad
\tilde{p}_3=\frac{p_3+V\omega_p}{\sqrt{1-V^2}},
\quad
\tilde{\omega}_p=\frac{\omega_p+Vp_3}{\sqrt{1-V^2}},
\eeq
here $\omega_p=\sqrt{p^2+M^2}$. In the limit $V\to 1$,
the momentum  $\tilde{p}_3$ is large,
$\tilde{p}_3\approx\tilde{\omega}_p\approx 2zP^2_N/{\cal M}_N$.
Wave functions of free fast-moving quarks become eigenfunctions
of polarization operator $\gamma^0\gamma^3$:
\beq
\bar{u}^{(\lambda)}(\tilde{p})=
P\sqrt{\frac{z}{m{\cal_M}_N}}u_0^{(\lambda) *}(1+\gamma^0\gamma^3),
\quad
\bar{v}^{(\lambda)}(\tilde{p})=
P\sqrt{\frac{z}{m{\cal_M}_N}}v_0^{(\lambda) *}(1+\gamma^0\gamma^3),
\la{tmp3}
\eeq
where the quark-antiquark spinors with spin up and down are:
\beq
u^\uparrow_0=\frac{1}{\sqrt{2}}\!
\left(\begin{array}{l}
1\\
0\\
-1\\
0\end{array}
\right)\! ,
\;
u^{\downarrow}_0
=\frac{1}{\sqrt{2}}\!\left(\begin{array}{l}
0\\
-1\\
0\\
1\end{array}
\right)\! ,
\;
v^{\uparrow}_0
=\frac{1}{\sqrt{2}}\!\left(\begin{array}{l}
1\\
0\\
1\\
0\end{array}
\right)\! ,
\;
v^{\downarrow}_0=\frac{1}{\sqrt{2}}\!\left(\begin{array}{l}
0\\
1\\
0\\
1\end{array}
\right)\! .
\la{tmp4}
\eeq
In fact, the matrix $\half(1+\gamma^0\gamma^3)$ is a projector
on the state with definite chirality.

Finally we get for the contribution of valence quarks to the one-quark
wave function
\beq
\bar{F}^{\lambda f}_{\rm lev}(z,\vec{p}_\perp)=
\tilde{f}^{f}_{\rm lev}
(\vec{p}_\perp,z{\cal M}_N-\varepsilon)
,
\la{level-1}
\eeq
where $\tilde{f}^{f}_{\rm lev}(\vec{p})$ is Fourier transform of the level
wave function. Using expression of \eq{level-wf} we obtain:
\beq
F^{f\lambda}_{\rm lev}(z,\vec{p}_\perp)=\left.\frac{\sqrt{{\cal
M}_N}\pi}{p^2}\left[
\left(p_3\tilde{j}(\vec{p})-p\tilde{h}(\vec{p})\right)\sigma_3+
\tilde{j}(\vec{p})\vec{p}_\perp\vec{\sigma}_\perp
\right]^f_{f^\prime}\varepsilon^{f^\prime\lambda}\!\!\right._{|\:{p_3=z
{\cal M}_N- \varepsilon}}
,
\la{level-final}
\eeq
where
\[
\tilde{h}(p)=\int_0^\infty\! dr r^2 h(r)\sqrt{\frac{2}{\pi}}\frac{\sin
pr}{r}, \quad
\tilde{j}(p)=\int_0^\infty \!dr r^2 j(r) \sqrt{\frac{2}{\pi}}\frac{\sin
pr -pr\cos pr}{pr^2}
\]

We see that, as it should be, the dependence on the nucleon momentum $P$
is cancelled in the final expression for the level contribution of
\eq{level-final} to the nucleon wave function. Also it is clear that
this contribution remains stable in the large-$N_c$ limit in the main
region of the quark momenta: $z_i\sim O(1/N_c)$, $p_\perp\sim O(1)$
(nucleon mass is proportional to the number of
colours $N_c$, factor $\sqrt{{\cal M}_N}$ is needed for the correct
normalization of the one-quark wave function).

Let us proceed now with the {\em sea quark contribution} to the wave
functions. According to \eq{1-quark} it can be expressed in terms of
the quark-antiquark pair wave function
$\Theta^{\lambda,\lambda^\prime}(\vec{p},\vec{p}^\prime)$ which, in
turn, is calculated through the finite-time Green function in the
external pion field.

The equal-time Green function is a sum of diagrams in the pion field.
These diagrams are similar to the ordinary Feynman graphs except for
the fact that the integration in the intermediate times in the graph is
going only in the interval $(0,T)$ (usually, one
integrates over the whole interval $(-\infty,\infty)$). It is easy to
show that these diagrams obey both the Dirac equation and necessary
boundary conditions at $t=0$ and $t=T$.

The diagrams represent the Green function as an expansion in the powers of
the pion field. One can rearrange this series in such a way that the
expansion is carried out in the powers of $\left( U^{\gamma_5}-1\right)$
instead. It can
be shown that the new expansion is in increasing powers of the gradients
of the pion field.

Let us restrict ourselves by the first non-trivial term of this
expansion. We call this approximation for the Green function an
{\em interpolation formula} \cite{DPP2,DPP3}. It becomes exact in
three limiting cases of the pion field: i) pion field is small
$\pi(x)\ll 1$, ii) pion field is slowly varying function of
coordinates with typical momenta $k_\pi\ll M$, iii) pion field momenta
are large  $k_\pi\gg M$. As a result, interpolation formula
usually works rather well in the problems related to nucleon-soliton:
its accuracy is typically around 10\%. \footnote{We checked the accuracy of
this formula, for example, in the calculation of the {\em nucleon
structure function}~\cite{Part1} confronting it to the exact
calculation accounting for all solutions of the Dirac continuum. This
problem is, of course, rather close to the calculation of wave
function. The interpolation formula was accurate enough.}

We introduce the Feynman Green function in the mixed $p$, $t$
representation:
\beq
{\cal G}(\vec{p},t)=\frac{\omega_p\gamma^0{\rm sign}(t)
-\vec{p}\vec{\gamma}}{2\omega_p}, \quad
\omega_p=\sqrt{\vec{p}^2+M^2}.
\eeq
The first-order correction to the pair wave function in the pion field
in the IMF is equal to
\[
\Theta_{\lambda_1,\lambda_2}(p_1,p_2)=\int\! d^3k\!
\sqrt{\frac{\omega_1\omega_2}{m^2}}
\delta\left(\frac{k}{\sqrt{1-V^2}}-(p_1)_z-(p_2)_z\right)
\delta\left(k_\perp-p_{1\perp}-p_{2\perp}\right)\times
\]\beq
\times u^{\lambda_1 *}(p_1)\gamma_0
{\cal G}^+(p_1)\frac{M\pi(k)\gamma_5}{\omega_1+\omega_2
-\frac{V\cdot
k}{\sqrt{1-V^2}}}
{\cal G}^-(-p_2)
v^{\lambda_2}(p_2),
\la{tmp2}
\eeq
signs $\pm$ label the sign of $t$ in the free Green functions.
Let us denote by $z_1$ and $z_2$ the fractions of the nucleon momenta
carried by the quark and antiquark correspondingly:
$(p_{1,2})_z=z_{1,2}P_N$. The energy denominator in \eq{tmp2} is small
only if $1>z_{1,2}>0$. Expanding it in the powers of nucleon momentum
and summing over spinor indices with help of \eq{tmp4}, we obtain:
\[
\Theta_{\lambda_1,\lambda_2}(z_1,z_2,p_{\perp 1},p_{\perp 2})=
\left(
M(z_1+z_2)(\sigma_3)_{\lambda_1\lambda_2}+
(z_2\vec{p}^\perp_1-z_1\vec{p}^\perp_2)
(\vec{\sigma}^\perp)_{\lambda_1\lambda_2}
\right)\times
\]\beq
\times
\frac{
{\cal M}_N M[\tilde{\pi}(\vec{p}_1+\vec{p}_2)]_
{p_{1z}=z_1{\cal M}_N,p_{2z}=z_2{\cal M}_N}
}{(z_1+z_2)z_1z_2{\cal M}_N^2+
(M^2+p_{\perp 1}^2)z_2+
(M^2+p_{\perp 2}^2)z_1},
\la{pair-pertubative}
\eeq
where $\sigma_i$ are Pauli matrices.\footnote{We normalize
\eq{pair-pertubative} in such a way that the pair wave
function is equal to
\[
\int\!\frac{dz_1d^2p_{1\perp}}{(2\pi)^2}
\int\!\frac{dz_2d^2p_{2\perp}}{(2\pi)^2}
a^{+(\lambda_1)}(z_1,p_{1\perp})
\ \Theta_{\lambda_1,\lambda_2}(z_1,z_2,p_{\perp 1},p_{\perp 2})
b^{+(\lambda_2)}(z_2,p_{2\perp}),
\]
where quark operators are normalized to
\[
\{
a^{+(\lambda_1)}(z_1,p_{1\perp})
a^{(\lambda_2)}(z_2,p_{2\perp})
\}=
\delta_{\lambda_1\lambda_2}\delta(z_1-z_2)(2\pi)^2
\delta^{(1)}(p_{1\perp}-p_{2\perp}),
\]\[
\{
b^{+(\lambda_1)}(z_1,p_{1\perp})
b^{(\lambda_2)}(z_2,p_{2\perp})
\}=
\delta_{\lambda_1\lambda_2}\delta(z_1-z_2)(2\pi)^2
\delta^{(2)}(p_{1\perp}-p_{2\perp}),
\]
}
As it should be, the nucleon momentum $P_N$ is cancelled out in this
expression. We see again that pair wave function has a finite limit
at large $N_c$ --- in the same sense as the one-quark wave function.

Collecting the terms in the perturbative expansion which correspond to
the expansion of $U^{\gamma_5}(x)$ in powers of $\pi$ we can
promote \eq{pair-pertubative} to the {\em interpolation formula}. It
has the following form:
\[
\Theta_{\lambda_1,\lambda_2}\!(z_1,z_2,p_{\perp 1},p_{\perp 2})\!=\!
\frac{M{\cal M}_N}
{(z_1\!+\!z_2)z_1z_2{\cal M}_N^2+
(M^2\!+\!p_{\perp 1}^2)z_2+
(M^2\!+\!p_{\perp 2}^2)z_1
}
\]
\[
\times
\left\{
\left(
M(z_1+z_2)(\sigma_3)_{\lambda_1\lambda_2}+
(z_2\vec{p}_{\perp 1}-z_1\vec{p}_{\perp 2})
(\vec{\sigma}^\perp)_{\lambda_1\lambda_2}
\right)
\tilde{\Pi}(\vec{p}_1+\vec{p}_2)-
\right.
\]\beq
\left.
-i\left(
M(z_2\!-\!z_1)\delta_{\lambda_1\lambda_2}+
i\varepsilon_{\alpha\beta}(z_1p^\perp_{2\alpha}\!-\!z_2p^\perp_{1\alpha})
(\sigma^\perp_\beta)_{\lambda_1\lambda_2}
\right)
\tilde{\Sigma}(\vec{p}_1\!+\!\vec{p}_2)]
\right\}_
{p_{iz}=z_i{\cal M}_N}
.
\la{pair-wf-fin}
\eeq
Here $\tilde{\Sigma}(\vec{k})$ is the Fourier transform of the
scalar component of $U^{\gamma_5}(x)$ and
$\tilde{\Pi}(\vec{k})$ is its axial component. In the soliton mean
field of \eq{hedgehog} they are equal to
\beq
\tilde{\Sigma}(\vec{k})=\int d^3x e^{-i\vec{k}\vec{x}}\left(\cos
P(r)-1 \right),
\quad
\tilde{\Pi}(\vec{k})=\int d^3x
e^{-i\vec{k}\vec{x}}(\vec{n}\vec{\tau})\sin P(r),
\eeq

The expression for the pair wave function of \eq{pair-wf-fin} has a clear
physical meaning.  The factor in front of the parenthesis in this
expression is, in fact, a light-cone wave function of the pion in the
leading order in $N_c$ calculated in Ref. \cite{pionwf}.\footnote{Pion
light-cone wave function of Ref.~\cite{pionwf}is obtained by substitution
$z_1+z_2=1$ (fractions should be measured relative to the momentum of
the pion) and integrating over transverse momenta
$p_{1\perp},p_{2\perp}$, with the condition $p_{1\perp}+p_{2\perp}=0$.
It corresponds to one definite ($\gamma_5$) component of the full wave
function. Also, in Ref.~\cite{pionwf} we took into account the dependence
of the dynamical quark mass on the quark virtuality which is neglected
here.} According to \eq{pair-wf-fin} one has to convolute this wave
function with a mean pion field in the nucleon in order to obtain
quark-antiquark pair wave function. This {\em factorization} is a
natural consequence of the large $N_c$ limit and strictly corresponds
to the soliton nature of the nucleon in this limit.  Moreover,
considering next terms in the expansion of the Green function one can
prove this factorization also in all orders.  For example, the second
term is the product of the two-pion light-cone wave functions
(calculated in our model in Ref.~\cite{maxchrist}) and the pion mean
field squared. At last, according to \eq{1-quark}, the sea
contribution to the one-quark wave function is a convolution of pair wave
function with the antiquark contribution to the light-cone wave function of
the discrete level. This formula is also guaranteed by the large $N_c$
limit.

Projecting the conjugated wave function of the level onto antiquarks we
obtain analogously to \eq{level-final}
\[
\bar{F}^{f,{\rm val}}_\lambda(\vec{p}_\perp,z)=-\sqrt{{\cal M}_N}
\bar{v}^{(\lambda)}(1+\gamma^0\gamma^3)f_{\rm lev}(
-\vec{p}_\perp,-z{\cal M}_N-\varepsilon)=
 \]\beq
=-\frac{\sqrt{{\cal M}_N}\pi}{p^2}\left[
(p_3j(p)-ph(p)(\sigma_1)^f_\lambda+j(p)
\left(p_x(\sigma_3)^f_\lambda+
ip_y\delta^f_\lambda
\right)\right].
\la{bar1quark}
\eeq
In order to obtain the sea contribution to the 1-quark wave function we
have to convolute the latter with the pair wave function of \eq{pair-wf-fin}:
\beq
F^{f,{\rm sea}}_\lambda(\vec{p}_\perp,z)=
-\frac{\sqrt{\cal M}_N\pi}{p^2}
\int\frac{dz^\prime d^2p^\prime_\perp}{2(\pi)^2}
\sqrt{\frac{z}{z^\prime}}\ \Theta_{\lambda\lambda^\prime}
(z,\vec{p}_\perp,z^\prime,\vec{p}^\prime_\perp)
\bar{F}^f_{\lambda^\prime}(z^\prime,\vec{p}^\prime_\perp)
.
\la{sea}
\eeq
Again this contribution does not depend on $N_c$ in the same manner as
the level wave function of \eq{pair-wf-fin}. The total one-quark light-cone
wave function is a sum of two contributions:
\beq
F^f_\lambda(\vec{p}_\perp,z)=F^{f,{\rm
val}}_\lambda(\vec{p}_\perp,z) +
F^{f,{\rm sea}}_\lambda(\vec{p}_\perp,z)
\la{onequark}
\eeq

\section{Distribution amplitudes of $N$ and $\Delta$}

Distribution amplitudes are, by definition, those
components of the Fock vector of state for a given particle, which
contain the lowest possible number of partons. For the nucleon
the distribution amplitudes are its three-quark wave function on the
light cone.  As it is well-known, distribution amplitudes describe
properties of the nucleon in hard exclusive processes.

The full wave function of the nucleon is given by
\eq{wf-total} and the wave function of the lowest component
(with $N_c$ quarks) is the product of $N_c$ wave functions
$F^f_\lambda(\vec{p}_\perp,z)$ of \eq{onequark}. However it is not
the end of the story. The point is that the minimum in \eq{hedgehog}
for the pion field, which corresponds to the nucleon, is degenerate. Any
pion field which is the flavour (or space) rotation or the translation
of the field of \eq{hedgehog} also gives a minimum of the action. In other
words, integration over the pion field in the functional integral has
{\em zero modes} which should be taken into account exactly.

It is well-known \cite{DPP2,Part1,Part2} that the integration over
translational zero modes leads to conservation of the momenta. In
the context of our calculation of nucleon distribution amplitudes, this
leads to the condition $z_1+\ldots z_{N_c}=1$ and
$p_{1\perp}+\ldots p_{N_c\perp}=1$. As to the
integration over flavour rotations, it gives rise to the quantum
numbers of the  nucleon-soliton: the state with given quantum numbers (in
$SU(2)$ flavour group they are: spin $J$, isospin $T$ and their
projections $J_3$ and $T_3$) is obtained as a projection of the
rotating soliton on the definite rotational wave function (which
appears to be Wigner D-function).\footnote{See, for example, Review
\cite{Rev1,Rev2} for the discussion of the rotational quantization of
baryon-soliton.}

As a result, accounting for rotational and translational zero modes, we
obtain the following expression for the nucleon wave function:
\[
\Phi^{J=T}_{J_3,T_3}(z_1,\ldots
z_{N_c},\vec{p}_{1\perp}\ldots\vec{p}_{N_c\perp})
=A\int\! dR\sqrt{2J+1}(-1)^{T+T_3}D^{(J=T)*}_{-T_3,J_3}(R)
\times
\]\[
\times\int\!\frac{dz_1d^2p_{1\perp}}{(2\pi)^2}\ldots
\int\!\frac{dz_{N_c}d^2p_{N_c\perp}}{(2\pi)^2}
\delta\left({\sum z_i-1}\right)
(2\pi)^2\delta^{(2)}\left(\sum\vec{p}_{i\perp}\right)
\times
\]\beq
\times R^{f_1}_{g_1}F^{g_1}_{\lambda_1}(\vec{p}_{1\perp},z_1)
\ldots
R^{f_{N_c}}_{g_{N_c}}F^{g_{N_c}}_{\lambda_{N_c}}(\vec{p}_{N_c\perp},
z_{N_c})
\la{3quark-general}
\eeq
Here $\Phi^{J=T}_{J_3,T_3}$ is the wave function for some
state from the rotational band of the soliton, $R$ is a rotational
matrix from the flavour group (we consider flavour group $SU(2)$ at the
moment), $A$ is the normalizing coefficient.

The integration over rotational and translational modes breaks down the
factorization of the nucleon wave function into the product of one-quark
wave functions. This is to be expected as rotation and translation are,
strictly speaking,  specific correction in $N_c$. Let us note that
the wave function of \eq{3quark-general} is symmetric under the exchange
of quarks, as it should be, since the colour wave function of this state
(omitted in this expression) is completely antisymmetric and, thus,
corresponds to the colourless baryon.

In the realistic applications one has to put $N_c=3$ and discuss states
with $J=T=1/2$ (nucleon) and $J=T=3/2$ ($\Delta$-resonance). The full
wave function of these states is given by
\eq{3quark-general}. However usually we are interested in the so-called
distribution amplitudes which are  the wave functions integrated over
all $p_\perp$:
\beq
\phi(z_1,z_2,z_3)=\delta(z_1+z_2+z_3-1)
\theta(z_1)\theta(z_2)\theta(z_2)\int\frac{d^2p_{1\perp}}{(2\pi)}
\ldots
\Phi^{J=T}_{J_3,T_3}(z_1,\ldots)
.
\la{def-distr-ampl}
\eeq
By definition, the normalization of the distribution amplitude is
chosen in such a way that the total integral over all three $z_i$ is equal
to unity.

It is known \cite{Dziembowski88,Boltz96} that relativistic invariance and
the symmetry considerations restrict the general form of the quark
distribution for the nucleon in such a way that it depends  on two
scalar functions only: completely symmetric $\phi_s(z_1,z_2,z_3)$ and
antisymmetric $\phi_s(z_1,z_2,z_3)$. For example, for the proton with
spin up we have:
\[
\phi^p_\uparrow(z_1,z_2,z_3)=\frac{\phi_s(z_1,z_2,z_3)}{\sqrt{6}}
\left(
2|u\uparrow d\downarrow u\uparrow>\!-\!
|u\uparrow u\downarrow d\uparrow>\!-\!
|d\uparrow u\downarrow u\uparrow>
\right)+
\]\beq
+\frac{\phi_a(z_1,z_2,z_3)}{\sqrt{2}}
\left(
|u\uparrow u\downarrow d\uparrow>-
|d\uparrow u\downarrow u\uparrow>
\right)
\la{8-def-wfn}
\eeq
Straitforward calculations based on \eq{3quark-general}, of course,
reproduce this general structure and give the concrete
expressions for the symmetric and antisymmetric parts of the
distribution amplitude.  However, these formulae are somewhat cumbersome
and we will not write them here.  As an example, let us give only the
expression for valence quark (level) conribution to the symmetric
and antisymmetric parts of the quark distribution:
\beq \phi_s(z_1,z_2,z_3)=c_0 \int\!
d^2x_\perp\left[2\tilde{g}(z_1,x_\perp) \tilde{g}(z_2,
x_\perp)\tilde{g}(z_3,x_\perp) \right.-\]
\[\left.  -
x_\perp^2\tilde{j}(z_2, x_\perp) \left(\tilde{g}(z_1,
x_\perp)\tilde{j}(z_3, x_\perp)+ \tilde{g}(z_3, x_\perp)\tilde{j}(z_1,
x_\perp) \right)\right]
,
\eeq\[
\phi_a(z_1,z_2,z_3)=\sqrt{3}c_0\int\!
d^2x_\perp x^2_\perp \tilde{j}(z_2, x_\perp) \left(\tilde{g}(z_1,
x_\perp)\tilde{j}(z_3, x_\perp)-
\right.
\]\beq
\left.
-\tilde{g}(z_3,x_\perp)\tilde{j}(z_1,
x_\perp) \right)
,
\la{nwf-final}
\eeq
where $\tilde{g}$ and $\tilde{j}$ are the Fourier transforms of the
level wave functions with $l=0$ and $l=1$ in the longitudinal
direction:
\[
\tilde{g}(z,x_\perp)=\int\!dx_3\exp\left(-ix_3(z{\cal M}_N-\varepsilon)
\right)\left[
h(r)-\frac{ix_3}{r}j(r)
\right],
\]\beq
\tilde{j}(z,x_\perp)=
\int\!dx_3\exp\left(-ix_3(z{\cal M}_N-\varepsilon)
\right)
\frac{ij(r)}{r}
\eeq
and $c_0$ is a normalizing constant.

In the non-relativistic limit, $j(r)\ll g(r)$ and only the symmetric wave
function $\phi_s(z)$ survives. The antisymmetric part is completely due to
the relativistic effects and sea quarks.

The total wave function should be calculated with the account for sea quarks
according to \eq{sea}. However, substituting the level and pair wave
functions into \eq{sea} we arrive at the integral which diverges at
small $z^\prime$ due to the phase volume factor $\sqrt{z/z^\prime}$.

In fact, we have already met the singularity of this type in the
calculations of pion wave function~\cite{pionwf} and generalized
distributions~\cite{Offw}.  The reason for this singularity is simple:
at small $z$ the quark virtuality
\beq
p^2_i\approx
\frac{M^2+\vec{p}_{i\perp}}{z_i}
\eeq
becomes large and we cannot
neglect any longer the dependence of the dynamical quark mass $M(p)$ on
this virtuality. The dependence of the constituent mass on the virtuality
is known in the instanton vacuum model \cite{DP1,DPP,Rev2};
substituting  $M(p)$ in \eq{sea} by this function we obtain the
converging sea quarks contribution which tends to zero at $z\to 0$.
As to the level contribution it remains non-zero at small $z$.

The total wave function accounting for the sea quarks can also be
represented in the form of \eq{nwf-final} where the functions
$\tilde{g}$ ¨ $\tilde{j}$ receive contributions both from valence and sea
quarks. Plots of the symmetric $\phi_s(z_1,z_2,z_3)$ and antisymmetric
$\phi_a(z_1,z_2,z_3)$ parts of the nucleon wave functions are displayed
in Fig.1 and Fig.2, correspondingly. It is seen
from these curves that the antisymmetric part of the wave function is
two orders of magnitude smaller than the symmetric one. This fact
is of a numerical origin, parametrically both parts of the wave function
are of the same order.

\begin{figure}
\centerline{
\epsfxsize=15cm
\epsfysize=15cm
\epsfbox{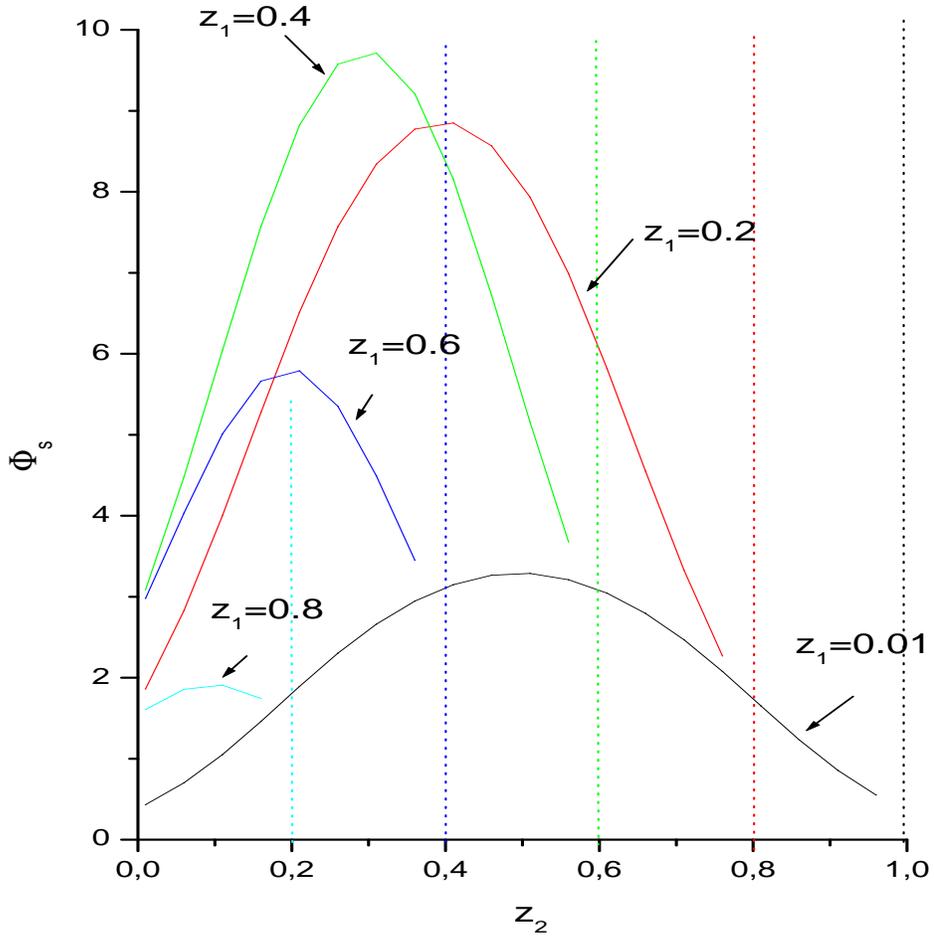}
}
\la{fig1}
\caption{Symmetric part of the nucleon quark distribution amplitude
$\phi_s(z_1,z_2,z_3)$ as a function of $z_2$ at different $z_1$.}
\end{figure}

\begin{figure}
\centerline{
\epsfxsize=15cm
\epsfysize=15cm
\epsfbox{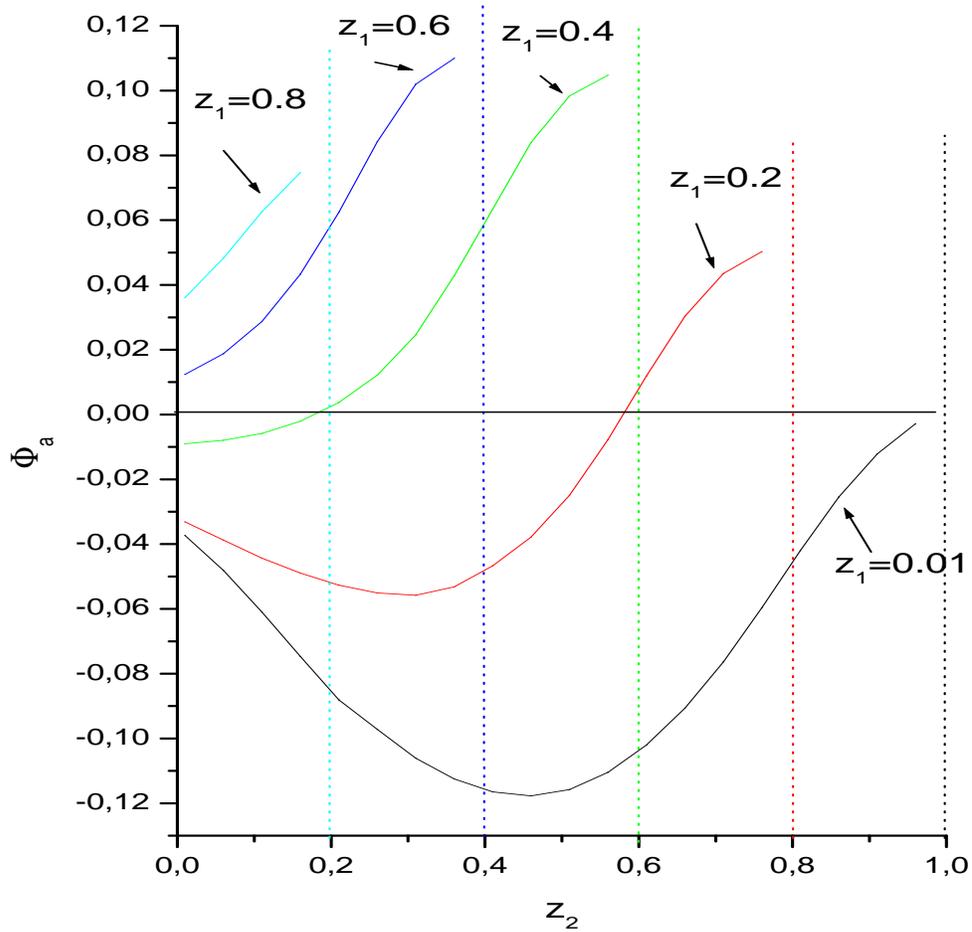}
}
\la{fig2}
\caption{Antisymmetric part of the nucleon quark distribution amplitude
$\phi_a(z_1,z_2,z_3)$ as a function of $z_2$ at different $z_1$.}
\end{figure}

In fact, the quark distribution amplitudes calculated here are valid only
in the region $zN_c\sim 1$. It is the main region for the nucleon wave
function. However, approximations used in the present paper break at
the end-points, at $z_i=0$ and $z_i=1$ but for different reasons.

In the $z\to 1$ limit  one quark is carrying almost all nucleon
momentum and the momenta of all other quarks are necessary small. This
limit contradicts the large $N_c$-approximation which leads to the
factorized wave function corresponding to independent quarks. The
$z\to 1$ asymptotics  corresponds to rather rare configuration of
the pion field and its contribution to the quark distribution is
exponentially small in $N_c$.\footnote{In principle, this situation
can be also described by semiclassical methods but one has to find a
new minimum in the functional integral over pion field corresponding to
the limit $z\to 1$. It can be shown that corrections to the mean pion
field become essential when parameter $1/N_c {\rm ln}(1-z)\ge
1$. Let us note that this situation is rather general and one faces
the same problem also in the calculation of structure functions at large
$N_c$~\cite{Part1,Part2}.}

Wave functions calculated here are neither valid in the limit $z\to
0$. The main reason is that at small $z$ the virtuality of nucleon
constituents become large and we are driven out of the region of the
applicability of chiral effective lagrangian of \eq{FI}. We have already
mentioned that we have to take into account dependence of the
dynamical mass of the quark $M(p)$ on its virtuality.
However it can be shown that this effect does not lead to nullification
of the quark distributions at $z=0$. It seems that more important is
an effect of gluons which are known to be produced intensively at small
$z$, at least in perturbation theory. As a consequence, the relative
contribution of pure 3-quark component decreases at small $z$. As it was
already mentioned, all effects of this kind were neglected here due to
the instanton vacuum parameter $(M\rho)^2\ll 1$.

The small antisymmetric part of the nucleon wave function contradicts
strongly to the commonly used parametrizations of quark distributions which
are obtained on the basis of form factor data or QCD sum rules (see, e.g.,
\cite{Boltz96}). Even more significantly our wave function differs from the
Chernyak-Zhitnizky's one \cite{Chernyak84a,Chernyak84b} which they suggested
many years ago analyzing QCD sum rules. Their wave function has
strong asymmetry $z_1\leftrightarrow z_2$ and even changes  sign as a
function of $z_1$ and $z_2$ (This looks strange for the wave function of the
nucleon which is the ground state in the sector with nonzero baryon
number). Our nucleon wave function is much closer to the so-called
asymptotic wave function (valid at arbitray high normalization point)
which is
\beq
\phi_s(z_1,z_2,z_3)=120\ z_1 z_2 z_3, \quad
\phi_a(z_1,z_2,z_3)=0;
\eeq
but still differs from it.

Data on the asymmetry of the nucleon wave function were reanalyzed
recently in Ref.~\cite{Braun2002} in the framework of the QCD sum rules
on the light-cone. According to the results of this analysis, the nucleon
wave function is not very far from the asymptotics. Let us also note
that the data on $\pi$-meson photoproduction on the
threshold~\cite{Pobylitsa01} also favor symmetric wave function.

Rotational symmetry of the nucleon-soliton allows one to calculate
immediately quark distributions for $\Delta$-resonance as well. For
$\Delta^+$-resonance with $J_3=+1/2$ we obtain
\[
\Phi^{\Delta^+}_{S_3=1/2}(z_1,z_2,z_3)=\frac{1}{\sqrt{3}}
\phi_\Delta(z_1,z_2,z_3)
\left(
|u\uparrow u\downarrow d\uparrow>+
|u\uparrow d\downarrow u\uparrow>-
\right.
\]\beq
\left.
-|d\uparrow u\downarrow u\uparrow>
\right)
\eeq
In other words $\Delta$-resonance is characterized only by one
symmetric wave function
\beq
\phi_\Delta(z_1,z_2,z_3)=\frac{c_0}{\sqrt{2}}\int\!
d^2x_\perp\left[\tilde{g}(z_1,x_\perp) \tilde{g}(z_2,
x_\perp)\tilde{g}(z_3,x_\perp) \right.
+
\]
\[\left.+
x_\perp^2\tilde{j}(z_2, x_\perp) \left(\tilde{g}(z_1,
x_\perp)\tilde{j}(z_3, x_\perp)+ \tilde{g}(z_3, x_\perp)\tilde{j}(z_1,
x_\perp) \right)\right]
\eeq

The main source of the experimental information on the nucleon wave
functions is the asymptotics of form factors at large $Q^2$. Four
form factors are measured: magnetic form factors of proton and neutron,
axial nucleon form factor and the transitional $\Delta\to N$ one.
In the region of hard $Q^2$, the process is factorized into a
product of the hard QCD part and corresponding wave function
(see, e.g.~\cite{Lepage79b,Lepage80} ). Asymptotics of the form factors
are:
\[
g_p\equiv Q^4G_{Mp}=2f\int\![dx][dy]
\left[
T_1\phi_s(x)\phi_s(y)
+x\leftrightarrow y\right]
,
\]\[
g_n\equiv Q^4G_{Mn}=-\frac{2f}{3}\int\![dx][dy]
\left[
(T_1-T_2)\phi_s(x)\phi_s(y)
+x\leftrightarrow y\right]
,
\]\[
g_A\equiv Q^4g_a(Q^2)=\frac{2f}{3}\int\![dx][dy]
\left[
(4T_1+T_2)\phi_s(x)\phi_s(y)
+x\leftrightarrow y\right]
,
\]\beq
g_{\Delta p}\equiv Q^4G_{Mp\Delta^+} =\frac{2\sqrt{2}f}{3}
\int\![d^3x][d^3y]
\left[
(T_1-T_2)\phi_s(x)\phi_\Delta(y)
+x\leftrightarrow y\right]
\la{as-formfactors}
\eeq
(we give here the simplified version of hard kernels in the assumption that
the antisymmetric part of the nucleon wave function is zero). Here
$[dx]$  denotes the integration over the fractions of total momentum
with the condition $x_1+x_2+x_3=1$, $f=(16\pi\alpha_s/9)^2$
and functions $T_1$ and $T_2$ are equal to
\[
T_1=\frac{1}{x_3(1-x_1)^2y_3(1-y_1)^2}+
\frac{1}{x_2(1-x_1)^2y_2(1-y_1)^2}-
\]\[
-\frac{1}{x_2x_3(1-x_3)y_2y_3(1-y_1)}
,
\]
\beq
T_2=\frac{1}{x_1x_3(1-x_1)y_1y_3(1-y_3)}
.
\eeq

Substituting our expressions for the wave functions into
\eq{as-formfactors} we see that integrals over $z_i$ are divergent due
to the end-point singularities both in the region $z_i=1$ and $z_i=0$,
where our approximations done are not valid (see above).
As to the divergence at $z_i=1$, it does not look serious: we know
anyway (from asymptotics of structure functions) that wave functions
should behave at least as $(1-x)^{3/2}$. The singularities are weak and
after any appropriate regularization the contribution of this region
to the form factors is negligible.

A discussion of the singularities at $x=0$ is much more involved. First,
we do not know any strict QCD theorem which states that the wave function
should be zero at $z=0$. The divergence is linear at $N_c=3$ but it
becomes worse at larger number of colours. As a result, the main
contribution  to the integral comes from this region.

Thus, the considered form factors are rather bad objects to be
treated in the framework of our model. We cannot calculate absolute value
of the form factors but can say only that they should be amplified by a
large parameter (connected in turn with parameter $(M\rho)^{-1}$). However
we can calculate {\em ratios} of form factors using the fact that divergent
part is universal for all form factors. In fact, this prediction is
based only on the limit of $N_c\to\infty$ and it does not use any
concrete realization of the nucleon-soliton model.

Ratios of the form factors in the large $N_c$ limit in comparison with
results of the two popular parametrizations of the nucleon wave function
are presented in Table~1. More extensive and complete analysis of
the phenomenological implications
of baryon distribution amplitudes can be
found in \cite{Stefanis:1992nw,Bergmann:eu,Stefanis:1992pi}. In
these works the corresponding analysis has been performed in the
framework of so-called heterotic approach.

We see that, qualitatively, large $N_c$ results agree with
data\footnote{Let us note that asymptotic wave function contradicts
the data strongly and is not displayed here for this reason.}.
Unfortunately the quality of the data is still rather bad.
For recent comprehensive reviews of theory and phenomenology of hard exclusive
reactions see Refs.~\cite{Sterman97,Stefanis:1999wy}.
\begin{table}
\begin{center}
\begin{tabular}{|c|c|c|c|c|}
\hline
Ratio& Ch.-Zh.~\cite{Chernyak84a} & G.-St.~\cite{Gari87}& Large $N_c$ &
Experiment \\
&&&(this paper)&\\
\hline $g_n/g_p$ & -0.48 & -0.1  & -1/3 &
$-0.45\pm0.1$~\cite{Rock92} \\ $g_A/g_p$  &  1.53 &  1.13 & 4/3  &
1.35~\cite{Carlson86a}\\ $g_{p\Delta}/g_p$ &0.01   &  0.81 &
$\sqrt{2}/12$ &$<0.3 \cite{Carlson86b}$\\
\hline
\end{tabular}
\end{center}
\caption{Form factor ratios.}
\end{table}

\section{Conclusions}
We have shown that the limit of large number of colours in QCD
allows one to tell much about the light-cone wave function of the nucleon.
 In this limit the nucleon is a soliton and its quark wave functions are
almost completely determined by the requirements of the large-$N_c$ factorization.
One needs to know only the wave function of the discrete level in the
self-consistent meson field and one-quark, two-quark, etc. wave
functions of the mesons which form this mean field. Then the
factorization, which is valid at $N_c\to\infty$, is enough to calculate
the whole light-cone wave function of the nucleon and, in particular,
its 3-quark component --- the distribution amplitude.

Put in such a way, this statement looks like a strict theorem. In the
present paper we realized this program for the concrete model of the
nucleon structure of \eq{FI} motivated by QCD instanton vacuum. The
main drawback of the model is, of course, the fact that the gluon
components of the wave function are parametrically suppressed
(by the parameter $(M\rho)^2\ll 1$) while we are used to think that
gluons play an essential role in the nucleon structure.

One should take into account that the distribution
amplitudes calculated in this paper are normalized at a very low
normalization point
($\mu\le 600$ MeV). Evolving these functions even to 1 GeV will lead to
significant amount of gluons in the wave function. Of course,
calculations of the gluon components of the wave functions (as
corrections in the density of instantons) are highly desirable. We plan
to discuss the role of gluons in the nucleon-soliton wave function in
the separate publication. For recent detailed works on evolution
equations for baryon distribution amplitudes see
Refs.~\cite{Braun:1999te,Bergmann:1999ud}.

However, we believe that even in the leading order, the large $N_c$ nucleon
wave function is interesting by itself. It was derived in
the relativistically invariant field theory and therefore it obeys all general theorems
and sum rules. It can be checked that the resulting wave function reproduces
correctly the structure functions of the nucleon calculated in
Ref.~\cite{Part1} and the generalized distributions of Ref.~\cite{Offw}.
This is to be expected as they are calculated in the same model, under
the same approximations. However, this demonstrates one more time that
the model is self-consistent. On the other hand it is known that the
structure functions obtained in Ref.~\cite{Part1} describe the
data, at least qualitatively.

The main features of the nucleon wave functions considered in this paper
are: i) relatively small contribution of the sea quarks to the component
with the lowest number of partons ii) soft distribution in the
transverse momenta of quarks with average value $\approx 400$ MeV
iii) the form which is much closer to the asymptotic wave function than
to the wave functions proposed on the basis of QCD sum rules \cite{
Chernyak84a}
(no zeroes or change of a sign) but which is still rather far from the
asymptotic form, iv) wave function is singular (does not vanish)
at
the end-points. These singularities can disappear as a result of
evolution or if one takes into account corrections in $N_c$.
Nevertheless it is natural to expect that factorization in the baryon
channel will appear at higher $Q^2$ than in meson sector. It is quite
probable that this is really the case in nature \cite{Braun2002}.

Let us note that the wave functions of mesons in the
instanton vacuum appears to be rather close to asymptotic one
in agreement
with the data (see Ref.~\cite{pionwf}). Baryon wave function
in the same model is far from the asymptotics and this fact again
corresponds to the data (asymptotics of the form factors). This is
natural, as in the limit $N_c\to\infty$ of the nature of mesons and
baryons is completely different. We conclude that this limit
describes experimental wave functions at least on the qualitative
level.

\vskip 0.5true cm
\noindent {\large\bf Acknowledgements} \\

\noindent
We are grateful to P.~Pobylitsa for numerous and very fruitful
discussions. The authors also acknowledge enlightening discussions with
V.~Braun, L.~Frankfurt,  A.~Ra\-dush\-kin, N.~Stefa\-nis, and M.~Strikman.
We are grateful to Klaus Goeke for hospitality at Bochum
University where part of this work has been done.

The work is supported by
Deutsche Forschungsgemeinschaft and Sofia Kovalevskaya Prize of
Alexander von Humboldt Foundation.
V.P. acknowledges partial support from the RFBR-0015-9606
grant.


\end{document}